\documentclass[12pt]{article}

\usepackage{graphicx}
\usepackage{booktabs}
\usepackage{xcolor}
\usepackage{colortbl}
\usepackage{subcaption}
\usepackage{amsfonts}
\usepackage{amsmath}
\usepackage{float}

\begin{document}
\title{Information and Contract Design 
for Repeated Interactions between Agents with Misaligned Incentives}

\author{
Nanda Kishore Sreenivas, Kate Larson\\
University of Waterloo \\
\{nksreenivas, kate.larson\}@uwaterloo.ca
}

\date{}

\maketitle

\begin{abstract}
    We study the consequences of information asymmetries and misaligned incentives  in settings with multiple independent agents. We model an interaction between a Sender, who holds vital private information but cannot act, and a Receiver, who must make decisions but is dependent on the Sender's information. We find that the Sender learns an optimal communication strategy that the Receiver reliably acts on. Importantly, this strategy is highly sensitive to the degree of conflict in the agents' rewards and the amount of environmental information the Receiver can already observe. We introduce a mechanism allowing the agents to form linear contracts, where a price is established for the information. We demonstrate that the Sender learns to use these payment structures to improve its rewards, though this comes at a cost of “fairness” between agents as the Sender is able to extract much of the Receiver’s surplus. This raises questions about fairness, contract design, and learning in the context of multi-agent systems.
\end{abstract}

\section{Introduction}

Agents are often faced with the challenge of making decisions under incomplete information. In many real-world scenarios, better-informed agents exist who can provide valuable information without direct intervention. If interests are fully aligned, the informed party can disclose everything it knows and the decision-maker’s action will deliver the best outcome for both. However, in practice, this alignment rarely holds. This misalignment introduces a strategic dimension to information sharing, where the information-rich agent may selectively disclose information to guide the decision-maker’s choices and steer outcomes toward its own objectives.

Online marketplaces exemplify this dynamic~\cite{Liu_2021_retail,zhang_2024}. While the platform possesses a global view of demand through aggregated data, the sellers control product selection and pricing. This creates a complex incentive structure: the platform seeks to steer seller behavior via strategic information disclosure to maximize commissions, yet its broader goals, like market-wide volume and variety, often clash with individual seller interests. This friction allows the platform to monetize its information advantage, offering paid recommendations that sellers purchase to refine their competitive strategy. Similar dynamics are also seen in ride-sharing and navigation apps.

In this paper we explore the tensions that arise in such settings. In particular, we propose a model where there is an information-rich \emph{Sender} agent and an action-taking, but less informed \emph{Receiver} agent. While both agents are cumulative-reward maximizers, their interests may be misaligned. While this type of interaction has been well studied in the one-shot setting through the Bayesian persuasion literature~\cite{kamenica_2011,kamenica_survey_2019}, our focus is on sequential decision-making, where the Receiver must take a series of actions over time without full access to the underlying state. The Sender cannot act directly but can influence the Receiver by sending signals. Before any observations are made, the Sender commits to a signalling policy---a probabilistic mapping from the true state of the world (captured by all payoff-relevant information) to the set of signals. The Receiver then uses this signal and its own partial observation to select actions. The rewards for both agents depend on the true state and the Receiver’s chosen action.

We further augment the model through the use of linear contracts. The Sender proposes a contract that stipulates the signalling policy along with the return share it seeks from the Receiver. If the contract is accepted, the agreed-upon fraction of the Receiver’s reward is transferred to the Sender. This allows the Sender to monetize its information advantage.

We conduct experiments to explore the behavior of the learning agents by varying (i) the degree of alignment between the agents’ incentives, and (ii) the quality of the Receiver’s own observations. Our key findings are as follows:
\begin{itemize}
\item The Sender learns effective signalling policies, and the Receiver learns to leverage these signals to improve its decision-making.
\item When incentives are misaligned, the influence of the signalling policy shifts depending on the quality of the Receiver’s private observations: it favors the Sender’s interests when the Receiver’s observations are poor, and vice versa.
\item When allowed to price its signalling policy through contracts, the Sender achieves higher cumulative rewards by learning to extract value from the Receiver. The pricing is also sensitive to the Receiver’s observation quality, with higher prices in low-information settings.
\end{itemize}

\subsection{Related Work}\label{sec:related_work}
%\kate{New related research below}
Our research incorporates three distinct, yet related fields, Bayesian persuasion, contract theory, and multi-agent reinforcement learning, to examine and understand how optimal communication and incentive alignment occurs between agents whose rewards are potentially misaligned.  In this section, we review key related research in these areas.

The foundational framework for our information design component is Bayesian Persuasion~\cite{kamenica_2011}, which models an informed Sender strategically influencing a Receiver's action based on information held by the Sender concerning the state of the world. Building upon this, recent reinforcement learning literature has explored dynamic Bayesian Persuasion, where interactions unfold in a sequential, Markovian manner. Key findings indicate that while optimal signalling is tractable for myopic Receivers, the problem becomes significantly more complex for far-sighted agents \cite{gan_2022,wu_2022}. Furthermore, research has also investigated signalling games where the Sender lacks commitment, resulting in coupled learning dynamics between the agents \cite{lin_2023}. Our work builds upon these dynamic settings by explicitly introducing reward misalignment, which fundamentally alters the Sender's optimal strategy and the informational efficiency of the system.

To address the challenge of reward misalignment, we integrate concepts from algorithmic contract theory, specifically by incorporating simple linear, payment-based contracts with the commitment-based information design of BP \cite{duetting_simple2019}. This allows the Sender to influence the Receiver's utility not only through reliable information transfer but also through direct financial incentives. This mechanism draws a parallel to the use of contracts in mixed-motive multiagent reinforcement learning (MARL) settings, where reward transfers are employed to resolve social dilemmas and enhance cooperation among otherwise self-interested agents \cite{christoffersen_2023_contracts}. Notably, the successful application of contracts in these MARL environments often relies on imposing structure on the large contract space, a methodological parallel we adopt to manage the complexity of the joint information and payment strategy space.

Our work distinguishes itself from the complementary research line concerning information markets~\cite{BergemannBonattiSmolin2018DesignPrice}. In information markets, an informed seller offers a menu of signalling policies and prices to a buyer, aiming to maximize revenue; the seller's utility is primarily revenue-driven. In contrast, our model maintains the core characteristic of Bayesian Persuasion: the Sender's utility is directly dependent on the Receiver's final action, not just the price paid for the information. This critical difference ensures that the design of the signalling policy and the contract is fundamentally driven by the operational outcome desired by the Sender, rather than solely by a pure profit motive.

Finally, our research provides insight for the wider field of MARL, particularly by contrasting with the  focus on cooperative settings within action-advising and communication literature \cite{dasilva_2019_survey}. While action-advising often involves an expert outside the environment providing advice to an agent \cite{dasilva_2017}, the research often, though not always (see~\cite{subramanian2022multi}),  assumes the advisor is well-intentioned or purely cooperative. By modelling agents with imperfectly aligned rewards, our work provides a necessary bridge to explore the complex dynamics of communication, learning, and incentive design in more realistic, mixed-motive MARL environments \cite{zhu2024survey}. Our findings offer new insights into how structured information and payments can enable constructive coordination even when agents harbour conflicting interests.

\section{Model}\label{sec:model}

In this section, we describe the components of our model and introduce the learning objectives.

We study settings where there are two different agents, a \emph{Sender (S)} and a \emph{Receiver (R)}, interacting in some environment. There are some key asymmetries in the problem. Namely,
\begin{enumerate}
    \item The Sender (S) has an informational advantage over the Receiver (R) in that it can observe more of the environment.  S can decide whether or not to share this information with R.
    \item The Receiver (R) has agency, in that it can act in the environment. The Sender (S) must rely on the actions of the Receiver.
    \item Both R and S obtain rewards from the environment. However, the rewards may be different and may not be fully aligned.
\end{enumerate}

We formalize the model by introducing two interrelated decision problems, one for the Receiver and one for the Sender, which can be viewed as a Markov Game. For ease of understanding, we present the model from the perspective of the two agents separately, coupling them when necessary.

\subsection{The Receiver}

Define $$\mathcal{M}_R=(S_R, A_R,P_R, \Omega_R, Re_R, \gamma_R)$$ to be the MDP for $R$, capturing the state space, the action space, the probability transition function, the observation space,  the reward function and discount factor respectively.

The Receiver does not observe the full environment state $s \in S_R$, but instead perceives a projection of it through an observation function $O: S_R \mapsto O_R$, where $O_R$ is the set of possible local observations available to the Receiver. The Receiver's effective observation space is $\Omega_R = O_R \times \Sigma$, where $\Sigma \supseteq A_R$ is a signal or message sent by the Sender, to the Receiver, specifying an action suggestion.

A policy for the Receiver, $\pi_R$, is a mapping from its observation space to (a distribution over) actions.
That is
$$
\pi_R:(O_R,\Sigma)\mapsto \Delta (A_R).
$$

The Receiver aims to maximize its expected discounted sum of rewards by following an optimal policy
$$
\pi_R^* = \arg\max_{\pi_R} \mathbb{E}\left[\sum_t \gamma^t_R Re_R(s_R^t, \pi_R(o^t, \sigma^t))\right]
$$
where $\sigma^t\in\Sigma$ is the signal sent by the Sender, and $o^t = O(s_R^t)$ is the Receiver's local environmental observation at time $t$.

\subsection{The Sender}
Define 
$$\mathcal{M}_S = (S_S, A_S, P_S, Re_S, \gamma_S)$$ to be the decision problem for Sender $S$, capturing the state space, action space, probability transition function, reward structure, and discount factor, respectively. 

We assume that $S_S=S_R\cup S_\emptyset$. That is, the state space of the Sender and Receiver is the same, except for one special state, $S_\emptyset$, which indicates that $S$ takes an action before $R$ considers any actions.  Furthermore, the environmental dynamics are the same as the Receiver, that is,  $P_S=P_R$, for all $s\in S_S\cap S_R$.  Recall that the Sender is not able to act directly in the same environment as the Receiver. Its actions, instead, take the form of action advice to the Receiver, $\sigma \in \Sigma$.

The Sender's reward function, $Re_S$ depends on the actions taken by the Receiver. In particular, 
$$Re_S: S_R\times A_R\mapsto \mathbb{R}$$ 
where $Re_S(s,a_R)$ is the reward earned by the Sender when the Receiver takes action $a_R$ in state $s$. Critically, even though the Sender can not take direct action in the environment of the Receiver, its reward is determined by how it influences the action choice of the Receiver, through the signals it sends.

A policy of the Sender is thus its signalling policy, which takes the form of a \emph{commitment probability}, $p$ with which the Sender commits to informing the Receiver of the best action to take, \emph{from the Receiver's perspective}. With probability $1-p$ the Sender will recommend an action that is optimal from its perspective. We further assume that the commitment probability is set at the start of an episode, before the Receiver agent takes any actions in the environment, \emph{i.e.}, the Sender's choice of $p$ is its action at the initial state $S_\emptyset$. We can formalize the policy of the Sender as follows:

\begin{displaymath}
    \pi_S(s) = 
\begin{cases}
    p \in [0,1] & \text{if } s = S_\emptyset \\
    \begin{cases}
        \pi_R'(s) & \text{with probability } p \\
        \pi_S'(s) & \text{with probability } 1 - p
    \end{cases} & \text{if } s \in S_R
\end{cases}
\end{displaymath}

Note that we assume the Sender has full observability of the environment in which the Receiver acts and, furthermore, is able to compute optimal policies for both the Receiver and itself.\footnote{We acknowledge that this is a strong assumption. We justify the assumption since it allows us to abstract away some of the learning problems of the Sender and focus on the central problem of information asymmetry and agency. Refer App.~\ref{apx:model} for a detailed explanation of modeling choices.}  
We define $\pi_S^{'}$ as the Sender's hypothetical optimal policy  \emph{if the Sender could act directly in the environment}. Further, we define $\pi_R^{'}$ as the hypothetical optimal policy from the Receiver's perspective, \emph{i.e.}, using the Receiver's reward structure. 
\begin{align*}
\pi_S^{'} = \arg\max_{\pi} \mathbb{E}\left[\sum_t \gamma^t_S Re_S(s_R^t, \pi(s_R^t))\right] \\
\pi_R^{'} = \arg \max_{\pi} \mathbb{E}\left[\sum_t \gamma^t_R Re_R(s_R^t, \pi(s_R^t))\right]
\end{align*}

Considering the Sender's policy, it is clear that the Sender's influence over the Receiver is through the choice of value of $p$, its commitment probability. The optimal policy of the Sender, $\pi_S^{*}$, is defined as
\begin{align*}
\pi_S^{*} = &\arg\max_{p\in[0,1]}\mathbb{E}[\sum_t \gamma_S^t (p Re_S(s^t, \pi_R^{*}(o^t,\pi_R^{'}(s^t)))\\
& +(1-p)Re_S(s^t, \pi_R^{*}(o^t, \pi_S^{'}(s^t)))) ].
\end{align*}
Critically, the optimal polices of both agents depend on each other, leading best-response dynamics. Determining whether the Receiver and Sender can learn to best respond to each other is a key focus of this work. 

This model allows us to answer the following questions:
\begin{description}
    \item[{\bf RQ1:}] Can the Sender learn optimal signalling policies ($p$) under different levels of misalignment? 
     Can the Receiver learn to better navigate the environment with the provided information?
    \item[{\bf RQ2:}] How does the quality of the Receiver's private observations ($O$) affect the agents' learned strategies and outcomes?
\end{description}

\subsubsection{Optimal Signalling}
\label{sec:optimal_p}
We derive the optimal signalling policy for the Sender. We denote the episodic return of an agent following policy $\pi$ as $J(\pi)$, given by:
\[
J(\pi) = \mathbb{E}_{\tau \sim \pi} \left[ \sum_{t=0}^{T-1} \gamma^t r_{t+1} \right],
\]

The expected utility of the Receiver over an episode is:
\begin{align}
U_R(p) = p \, V_{RR} + (1-p) \, V_{RS},
\end{align}
where $V_{RR} = J_R(\pi'_R)$ and $V_{RS} = J_R(\pi'_S)$.

Let $U_R^0$ denote the Receiver's independent episodic reward, which primarily depends on its quality of observations. Then, the obedience constraint dictates that:
\begin{align}
U_R(p) \ge U_R^{0} \implies p \, V_{RR} + (1-p) \, V_{RS} \ge U_R^{0}.
\end{align}

\begin{align}
p_{\min} = \frac{U_R^0 - V_{RS}}{V_{RR} - V_{RS}}.
\end{align}

$U_S(p)$ is linear and decreasing in $p$ (because $V_{SS} \ge V_{SR}$ by definition), so the optimal signalling policy is:
\begin{align}
p^* = \min(\max(p_{\min}, 0),1)
\label{eq:optimal_p}
\end{align}

Note that $V_{RS}$ and $V_{RR}$ depend on the degree of misalignment between $S$ and $R$ while $U_R^0$ varies with the level of Receiver's observability.

\subsection{Contracts and Information Pricing}

We introduce the possibility of the Sender charging for information through the use of ``take it or leave it'' linear contracts~\cite{duetting_simple2019}. 
A contract is an agreement between the Sender and the Receiver where the Receiver agrees to share a portion of their collected reward in exchange for information provided by the Sender. While contracts can take many forms, linear contracts, captured by a single parameter, $c\in[0,1]$, have been shown to be very powerful and robust in settings with uncertainty~\cite{duetting_simple2019,carroll2015robust}.
This gives rise to several questions:
\begin{description}
    \item[{\bf RQ3:}] Can the Sender learn how to use contracts effectively? 
    \item[{\bf RQ4:}] How does the use of contracts impact the Receiver's accumulated reward?
    \end{description}

To explore these questions we expand the action space  of the Sender and Receiver.
The Sender announces $\langle p, c\rangle$ to the Receiver, specifying its signalling policy ($p$) and the reward share $c\in[0,1]$ it will collect from the Receiver's collected rewards. The Receiver can decide to accept or reject the proposal. If the proposal is rejected, the Receiver must act in the environment with no further interaction from the Sender. If the proposal is accepted, the process is the same as described earlier, except that the reward structure changes. The effective reward structures become
\begin{align}
    Re_S^{'} = Re_S + c Re_R \\
    Re_R^{'} = (1-c)Re_R     
\end{align}
%The policies for both the Sender and the Receiver are adapted to account for this modification of reward structure. 

The optimal contract is characterized by:
\begin{equation}
p^* =
\begin{cases}
p_{\min}, & c < \hat{c},\\
1, & c > \hat{c}.
\end{cases}
\label{eq:optimal_pc}
\end{equation}

where $\hat{c}$ is a threshold that depends on the degree of incentive alignment (derived in App.~\ref{apx:contracts}). For $c < \hat{c}$, transfers are weak and the agents' incentives remain misaligned. So, the interaction reduces to the signalling setting ($p^* = p_{min}$), and there is little surplus to be extracted through contracts. For $c > \hat{c}$, transfers are large enough to align incentives, and full revelation $p^* = 1$ becomes optimal.

\section{Experiments}

In this section, we present our experimental findings. We ground our work in two settings. The first is a classic recommendation letter scenario from the Bayesian persuasion literature~\cite{dughmi_survey_2017}, while the second is a grid-world environment which allows us to explore the impact that reward alignment has on agents' learned policies.

\subsection{Recommendation Letter} \label{sec:rec_letter}
In the recommendation letter problem, there are two agents, a professor (Sender) and a recruiter (Receiver)~\cite{dughmi_survey_2017}. The professor is writing a recommendation letter for their student who is being recruited by the recruiter. The student is either a strong candidate or a weak candidate, and the student quality is known to the professor but not to the recruiter. The recommendation letter serves as a binary signal (recommend/don't recommend) from the professor to the recruiter. The recruiter receives a reward of +1 for hiring a strong student and -1 for hiring a weak student. If no student is hired, both agents receive a reward of 0. The professor receives a reward of +1 whenever their student is hired, regardless of quality.
This problem captures the challenges of asymmetric information and misaligned incentives. If the professor (Sender) truthfully reported student quality, the recruiter (Receiver) would only hire strong students. By recommending all strong students and randomly recommending weak students, the professor can increase their expected utility.

Formally, a student’s type is drawn from $\{S,W\}$ with prior $Pr(S) = p_0 \leq \tfrac12$

The professor chooses a signalling rule $\sigma:\{S,W\}\to \Delta(\{R,\bar R\})$, parameterized by
\[
p_1 = \Pr(R|S), \qquad p_2 = \Pr(R|W).
\]

We defer details to Appendix~\ref{apx:letter}, but the optimal signalling policy for the professor is given by:
\[
p_1^* = 1, \qquad 
p_2^* = \frac{p_0}{1-p_0}
\]

Consider a specific example, where $p_0 = 1/3$, i.e, a randomly drawn student is strong with probability $\tfrac13$. Without any information about a particular student, the expected utility of the recruiter for hiring a student is negative ($-\tfrac13$), and, therefore, the recruiter will not hire any student. Under perfect information, i.e., the professor recommends a strong student and does not recommend a weak student, the recruiter will hire only the strong students netting an expected utility of $\tfrac13$ for both parties. However, if the professor uses the optimal signalling policy, i.e., recommend a weak student with one half probability, then the professor increases their expected utility to $\tfrac23$. 
We arrive at the same signaling policy $p^*=\tfrac{1}{2}$ using our analysis from Section~\ref{sec:optimal_p} (see App.~\ref{apx:letter} for details).

\subsubsection{Experimental Setup}
We model the learning processes of both agents using multi-armed bandits. The Sender's policy is a tuple $\langle p_1, p_2 \rangle$ where $p_1$ is the probability that the Sender provides a good recommendation if the student is strong ($P(R|S)$), while $p_2$ is the probability that the Sender provides a good recommendation if the student is weak ($P(R|W)$). Thus, the arms for the Sender's bandit problem correspond to different signalling policies. 
The Receiver observes the signalling policy of the Sender and the recommendation. This forms the context for a contextual bandit problem with two arms, with one arm corresponding to the hire decision and the other arm corresponding to the not hire decision. Rewards for both the Receiver and Sender are observed after the hire/not hire decision and arm-values are updated.

%We ensure that there is a finite number of arms for the Sender's bandit problem by discretizing $p_1$ and $p_2$ into $0.05$ increments
We ensure there is a finite number of arms for the Sender's bandit problem by discretizing $p_1$ to the set $\{0,0.5,1\}$ and $p_2$ into $0.05$ increments.~\footnote{We also experimented with finer discretization and observed similar results; see Figure~\ref{fig:letter_heatmap_fullrange} in App.~\ref{apx:letter_res}} Each trial consists of 200,000 interactions, and we define an episode to be 50 interactions. At the start of an episode, the Sender commits to a fixed strategy $\langle p_1, p_2\rangle$. The underlying learning algorithm was a discounted-UCB algorithm \cite{garivier_2011_bandits}.~\footnote{We use a discounting rate of $0.95$.} All of our results are averages computed over 100 trials.

\subsubsection{Recommendation Letter Results}

\begin{figure}[!h]
    \centering
    \begin{subfigure}{0.99\columnwidth}
        \centering
        \includegraphics[width=\linewidth]{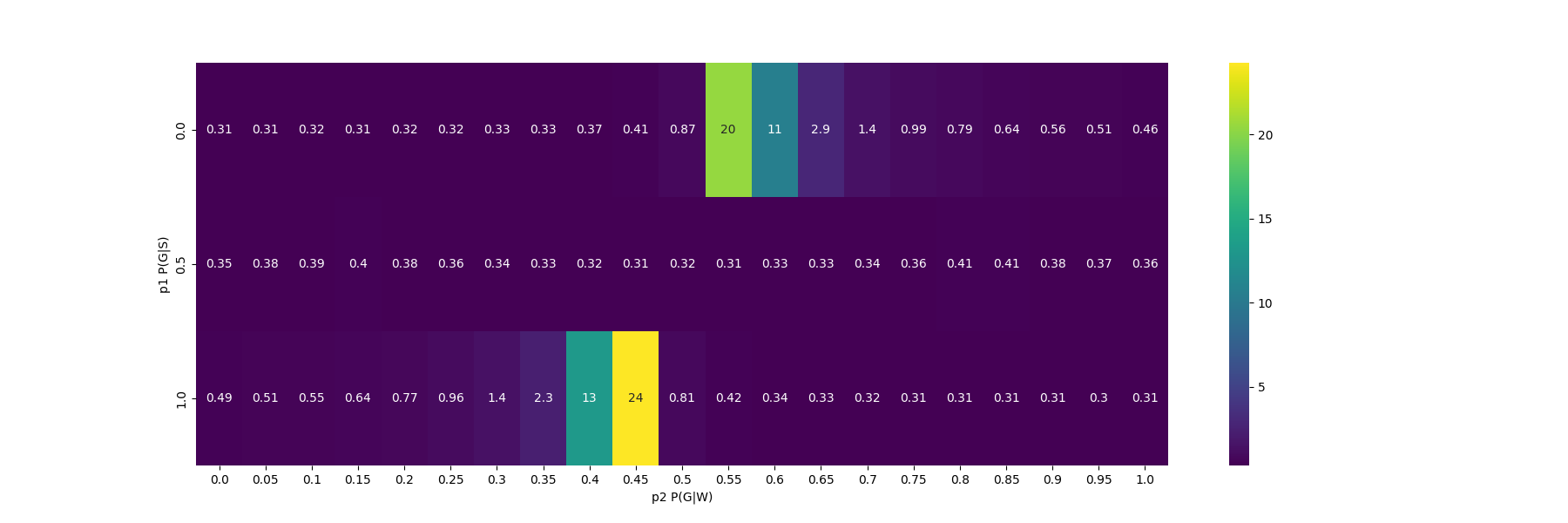}
        \caption{Average selection percentages of signalling strategies $\langle p_1, p_2 \rangle$}
        \label{fig:letter_infodesign}
    \end{subfigure}
    \begin{subfigure}{0.95\columnwidth}
        \centering
        \includegraphics[width=\linewidth]{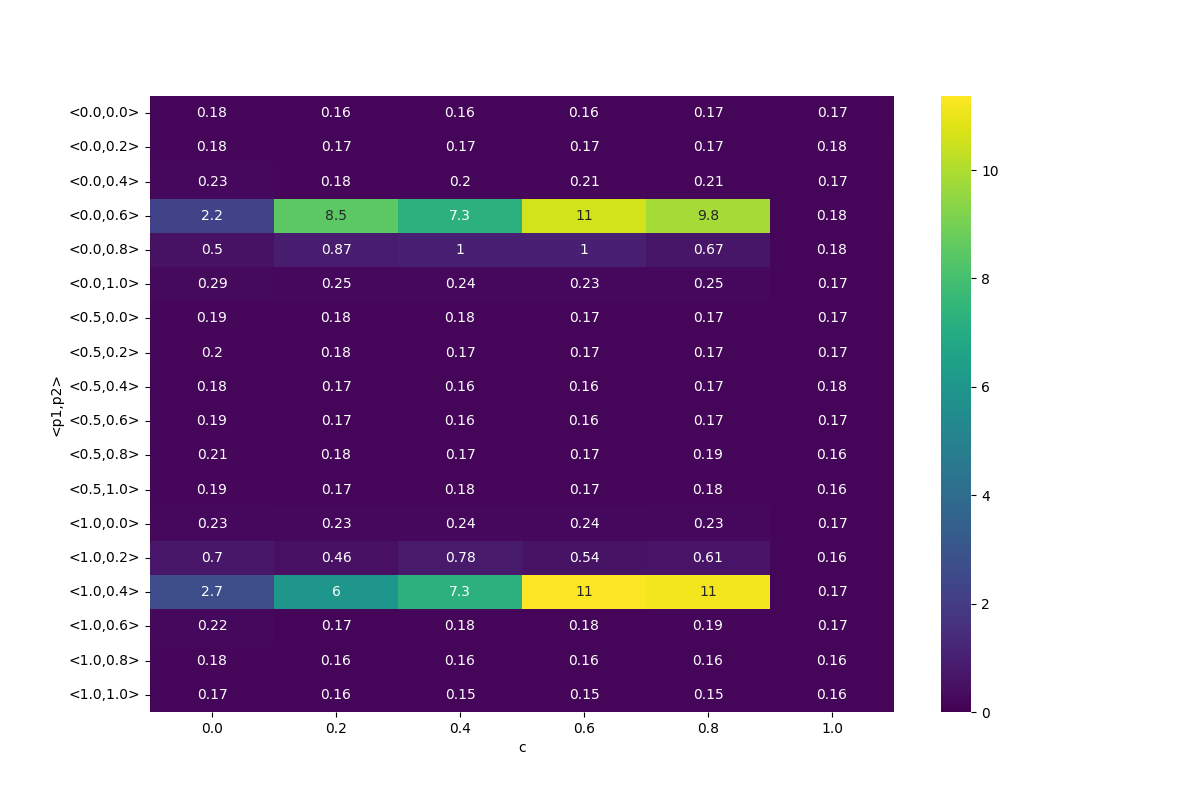}
        \caption{Average selection percentages of contract proposals}
        \label{fig:letter_contract}
    \end{subfigure}
    \caption{Results for Recommendation Letter}
\end{figure}

We first determine whether agents can learn optimal policies for the recommendation letter problem. We instantiate an instance of the problem where the prior probability that a student is strong is $\frac{1}{3}$.

We first study the case where the Sender learns a signalling policy. The results are shown in Figure~\ref{fig:letter_infodesign}. 
In particular, we notice that the Sender quickly settles on two signalling policies, $\langle 1.0, 0.45\rangle$ and $\langle 0.0, 0.55 \rangle$, resulting in average rewards of $0.54$ for the Sender (professor) and $0.06$ for the Receiver (recruiter). 
We observe that the average rewards are close to the theoretical optimal rewards, and that signalling strategy $\langle 1.0, 0.45\rangle$ is a close approximation to the optimal strategy.
This experiment answers \textbf{RQ1}, i.e., the Sender (professor here) learns the optimal signalling policy and the Receiver (recruiter) learns to best respond to it.

In our second set of experiments we studied the impact of the addition of contracts to answer \textbf{RQ3}. The Sender's strategy is enriched to be a vector $\langle p_1, p_2, c\rangle$ where contract $c\in[0,1]$ is the fraction of the Receiver's reward that is paid to the Sender if the contract is accepted.  In our experiments, we use the same discretization for $p_1$ and $p_2$ as before and discretize $c$ in 0.2 increments, resulting in a 108-arm bandit problem. The Receiver's problem is the same as before, but with an enlarged context (the signalling strategy and the proposed contract), but with the caveat that if the contract is rejected, the Sender sends no signal as to the strength of the student and so the Receiver must make a decision (hire/don't hire) without information.

Figure~\ref{fig:letter_contract} shows the overall contract proposals made by the Sender, while Figures~\ref{fig:overall_contract_acceptance} and~\ref{fig:contract_acceptance_rates} in App.~\ref{apx:letter_res} present the contract-specific acceptance rates. We first observe that the signalling strategy of the Sender quickly converges to the optimal signalling strategies we observed before, but there is more variability around the contract price. While we see that the use of contracts does increase the Sender's average utility to 0.55 while dropping the Receiver's utility to 0.02, we hypothesize that the benefit of contracts is small in this context since there is little surplus to extract from the Receiver.

We note that this is in line with our theoretical analysis. Substituting the expected utility values from this setting into~\eqref{eq:c_hat}, we obtain $\hat{c} = 1$. Thus, the optimal information contract~\eqref{eq:optimal_pc} is to implement the same optimal signaling policy as in the non-contract case while charging a cost $c < \hat{c} = 1$.

\subsection{Gridworld Experiments}

We now explore the possibility of learning signalling policies and contracts in a more complex setting, where we can control both the reward alignment and information asymmetry between the Sender and Receiver.

\begin{figure}[]
    \centering
    \begin{tabular}{|m{0.5cm}<{\centering}|m{0.5cm}<{\centering}|m{0.5cm}<{\centering}|m{0.5cm}|}
        \hline 
        \includegraphics[width=1.5em]{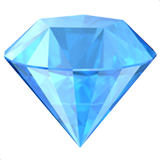} & & & \\[0.5cm] \hline
         & \cellcolor{cyan!20} & \cellcolor{cyan!20} & \cellcolor{cyan!20} \\[0.5cm] \hline
         & \cellcolor{cyan!20} & \includegraphics[height=1.5em]{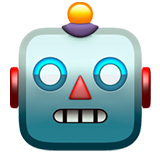} & \cellcolor{cyan!20} \\[0.5cm] \hline
        \includegraphics[height=1.5em]{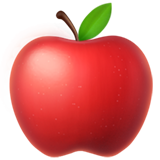} & \cellcolor{cyan!20} & \cellcolor{cyan!20} & \cellcolor{cyan!20} \\[0.5cm] \hline
    \end{tabular}
    \caption{The full environment is a $10\times10$ grid; for clarity, we display a $4\times4$ subsection illustrating the key elements. The robot marks the Receiver and objects are shown on the map. The Sender observes the full grid, while the Receiver sees only the highlighted cells (for visibility $v=1$).}
    \label{fig:gridworld}
\end{figure}

\subsubsection{Environment and Reward Structures}
Our environment is shown in Figure~\ref{fig:gridworld}. It is a simple 10 by 10 grid world with two types of objects: apples and diamonds. The Sender can observe the entire grid,  but can not move in the environment. The environment is initialized by placing the Receiver at a cell chosen uniformly at random, and placing one apple and one diamond randomly in the remaining cells. As the objects are collected, they spawn in a new location. The Receiver is able to move and collect objects but has limited observability. 
To answer \textbf{RQ2} we control the observations of the Receiver through a visibility parameter $v$. This parameter controls  the observation radius of the Receiver in terms of the Moore neighbourhood around the Receiver, i.e., the Receiver observes all cells within $v$ steps in each direction. %To answer \textbf{RQ2} experimentally
In particular, we consider two scenarios to control for the quality of Receiver observations --- low-visibility scenario ($v=1$, average observability near 10\%), and high-visibility scenario ($v=5$, average observability near 50\% of the grid).

While the Receiver can collect both apple and diamond objects, we structure the rewards of the agents so that their interests are potentially misaligned. In particular, the reward functions of the agents are a vector $\langle r_a, r_d\rangle$ where $r_a$ is the reward an agent receives for a collected apple while $r_d$ is the reward per collected diamond.  We set the reward vector for the Receiver agent, $r_R$ to be $\langle 1, 0\rangle$ (i.e. it only cares about collecting apples). We capture the degree of misalignment between the Receiver and the Sender by a parameter $\theta$, the angle between two reward vectors, and set the reward vector of the Sender agent to be $r_S = \langle \cos\theta, \sin\theta\rangle$. Thus, fully aligned agents ($\theta = 0$) have the same reward vectors while fully misaligned agents ($\theta=180$) have reward vectors $\langle 1, 0\rangle$ and $\langle -1, 0 \rangle$. 
We consider $\theta \in \{0^\circ, 30^\circ, 45^\circ, 60^\circ, 90^\circ, 180^\circ\}$ (corresponding reward vectors in App.~\ref{apx:gridworld}), covering aligned, partially divergent, orthogonal, and diametrically opposed objectives.
These experiments address \textbf{RQ1}: whether the Sender learns optimal signalling policies and how these policies change under varying degrees of reward misalignment.

\begin{figure}[ht]
    \includegraphics[width=0.95\textwidth]{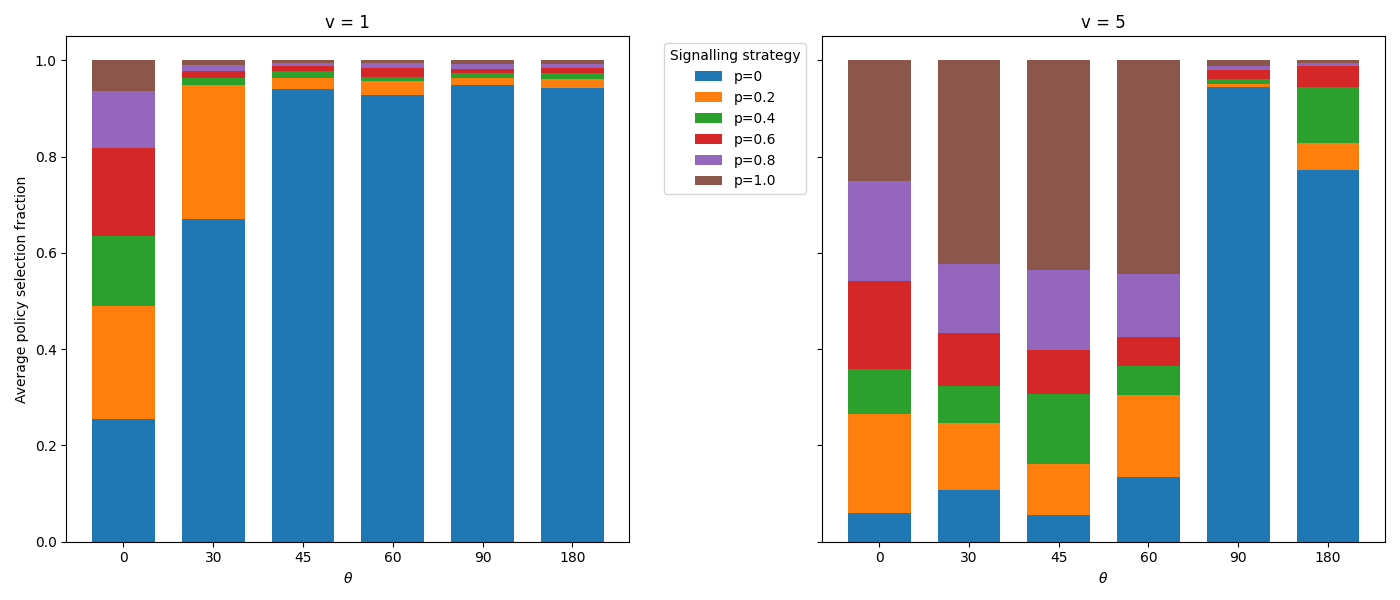}  
    \caption{Gridworld: Average selection frequencies of each signaling strategy for different reward alignment angles ($\theta$). Left: Receiver with low visibility ($v=1$). Right: high visibility ($v=5$).}
    \label{fig:compressed_arm_selection_nocost}
\end{figure}

\subsubsection{Learning Processes}
As described in Section 2, we assume that the Sender can compute optimal policies for moving in the grid world from its own perspective ($\pi_S^{'}$) and from the perspective of the Receiver($\pi_R^{'}$). It uses these policies to make action recommendations to the Receiver. Given these policies, we are interested in understanding what signalling and contract policies the Sender  learns, and how the Receiver learns how to respond. As in the recommendation letter example, we use bandits as the underlying learning mechanism for the Sender.  We consider two settings: the signalling setting without payments and the contract setting as described in Section 2.3. The Sender's policy takes the form of a tuple $\langle p, c\rangle$, $p,c\in[0,1]$, where $p$ is the probability that the Sender recommends the action according to $\pi_R^{'}$ (and with probability $1-p$ it recommends the best action from its perspective, according to $\pi_S^{'}$). Parameter $c$ is the cost, which specifies what fraction of the reward collected by the Receiver should be shared with the Sender. For example, if $p=1$ and $c=0$ then the Sender always sends optimal action information for the Receiver and asks for no compensation, while if $p=0$ and $c=1$ then the Sender always recommends the best action for itself and demands all the Receiver's rewards. We discretize the strategy space into $\{0.0, 0.2, \ldots, 0.8, 1.0\}^2$, resulting in 36 arms. After an arm is selected, the arm's value is updated with the Sender's episodic reward. We use discounted $\epsilon$-greedy with discount factor $0.9$ and an $\epsilon$ schedule decaying linearly from $1$ to $0.05$ over the first $75\%$ of training.

The learning problem of the Receiver is more complicated since it must learn whether to accept or reject the contract and, if the contract is accepted, whether to accept the action recommendation or act on its own.
We use tabular Q-learning to learn whether or not to accept a contract, where the state space is all possible Sender proposals $\langle p,c \rangle$, the action space is to accept or reject the contract, and the Receiver's episodic reward is used for updates.\footnote{We use a discounting rate of 0.9, a learning rate of 0.1, and an exploration constant of 0.05.} We use PPO from Stable-Baselines3 to learn whether the Receiver should follow the Sender's recommended action (with $n_{steps}=256$, batch size $64$, $n_{epochs}=10$). 
If the action recommendation is not followed, then the Receiver follows a simple heuristic strategy that greedily moves towards the closest observed apple or takes an action at random if no apples are observable. \footnote{We also experimented with using BFS as the default Receiver policy. The results, shown in Tables~\ref{tab:nocost_rewards_bfs} and~\ref{tab:cost_rewards_bfs} in App.~\ref{apx:gridworld}, follow similar trends to those reported here.} Agents are trained for 2000 episodes and evaluated on 100 test episodes. All reported values represent averages across 10 independent trials, each using a different random seed.

%\noindent{\bf Signalling Strategies:}

\begin{table}[ht]
\centering
%\scriptsize
\begin{tabular}{ccccc}
\hline
$\theta$ & Receiver & Sender & Apple & Diamond \\ \hline
\multicolumn{5}{c}{v = 1} \\
\hline
0   & 53.72 (0.48) & 53.72 (0.48) & 74.79 (0.46) & 3.88 (0.15) \\
30  & 45.24 (0.66) & 50.31 (0.87) & 65.57 (0.73) & 27.71 (2.51) \\
45  & 33.24 (0.49) & 49.54 (0.51) & 53.32 (0.49) & 45.13 (0.74) \\
60  & 11.00 (0.69) & 49.35 (0.82) & 31.32 (0.70) & 62.36 (1.15) \\
90  & -6.26 (1.31) & 47.67 (2.51) & 14.59 (1.23) & 68.50 (2.38) \\
180 & -6.22 (1.09) & -35.38 (1.06) & 14.63 (1.07) & 68.33 (2.00) \\
\hline
\multicolumn{5}{c}{v = 5} \\
\hline
0   & 53.67 (0.54) & 53.67 (0.54) & 74.74 (0.52) & 3.91 (0.18) \\
30  & 48.64 (2.26) & 43.05 (1.51) & 69.78 (2.24) & 7.49 (2.62) \\
45  & 46.59 (3.41) & 32.20 (1.62) & 67.81 (3.37) & 7.72 (3.35) \\
60  & 43.29 (4.35) & 18.16 (1.46) & 64.64 (4.26) & 8.27 (3.01) \\
90  & 35.34 (2.49) & -3.42 (1.55) & 56.62 (2.33) & 17.82 (1.39) \\
180 & 31.83 (3.50) & -74.83 (2.96) & 53.35 (3.23) & 16.45 (2.81) \\
\hline
\end{tabular}

\caption{Signalling Setting: Average episodic rewards and the mean number of apples and diamonds collected (SD in parentheses).}
\label{tab:nocost_rewards}
%From file: narval_logs/log_1730828243.9544244.json
\end{table}

\subsubsection{Signalling Strategies}
First, we look at the case where  the Sender does not charge a price for information.  The average rewards for both agents and the number of objects collected on average in an episode are shown in Table~\ref{tab:nocost_rewards}. When the angle between their reward vectors, $\theta$ is 0, they are fully aligned, and therefore, they are interested in apples only and receive the same reward.  As the degree of reward misalignment increases (i.e., as the Sender’s preference shifts toward diamonds while the Receiver continues to prefer apples), the outcomes depend strongly on the Receiver’s visibility. Under low visibility ($v = 1$), the number of collected diamonds rises with $\theta$, reflecting the Sender’s increasing influence on the Receiver’s actions. But, under high visibility ($v = 5$), the Receiver collects more apples, as improved observability allows it to rely more on its own policy rather than the Sender’s signals. This shift in outcomes is also reflected in the Sender’s learned signalling policy (see Figure~\ref{fig:compressed_arm_selection_nocost}). When the Receiver’s observability is low, the Sender typically adopts signalling strategies with $p = 0.0$, meaning that it consistently recommends actions aligned with its own interests rather than the Receiver’s. As visibility increases, the Sender’s learned strategy uses higher $p$ values (until $\theta < 90$). The exact value of $p$ depends on the degree of reward alignment: the more misaligned the agents, the lower the likelihood of Sender's advice being sampled from the Receiver-optimal policy. 

As a baseline, we consider the case where the Receiver navigates the environment without any interaction with the Sender. In the low visibility scenario ($v=1$), its average episodic reward (over 100 episodes) is -20.81, whereas, in the high visibility scenario, it increases to 11.69. This result addresses \textbf{RQ1}, showing that the Sender learns good signalling policies and the Receiver learns to use the signals to improve its rewards. These signalling policies are sensitive to the degree of reward misalignment. This experiment also contributes to answering \textbf{RQ2}, showing that the Sender’s signalling strategies tend to align more closely with the Receiver’s interests when the Receiver’s private observations are of higher quality.

\subsubsection{Contract Strategies}

We now explore whether the Sender learns to effectively use contracts to price the information sent to the Receiver. The average episodic rewards and the mean number of collected objects per episode are presented in Table~\ref{tab:cost_rewards}. The contract strategies learned by the Sender are illustrated in Figure~\ref{fig:heatmaps} in App~\ref{apx:gridworld}, where each heatmap summarizes the relative frequency of different contract strategies for a given value of reward misalignment ($\theta$) and Receiver observability ($v$). In each heatmap, the rows represent the values of $p$, the commitment probability and the columns represent the values of $c$, the cost.

\begin{table}[t]
    \centering
    %\scriptsize
    \begin{tabular}{ccccc}
        \hline
        $\theta$ & Receiver & Sender & Apple & Diamond \\
        \hline
        \multicolumn{5}{c}{v=1} \\ \hline
        0 & 6.31 (3.66) & 78.49 (22.90) & 64.04 (11.39) & 3.22 (0.37) \\
        30 & 3.44 (4.85) & 83.07 (8.40) & 64.95 (4.72) & 14.71 (6.12) \\
        45 & 5.52 (3.61) & 69.26 (6.90) & 58.77 (7.67) & 23.01 (11.29) \\
        60 & 3.76 (3.70) & 50.46 (6.50) & 46.23 (12.06) & 31.30 (15.33) \\
        90 & -3.42 (3.20) & 34.78 (8.89) & 24.63 (12.13) & 49.32 (16.81) \\
        180 & -8.37 (0.91) & -33.91 (0.81) & 13.13 (0.97) & 63.07 (2.70) \\ \hline
        \multicolumn{5}{c}{v=5} \\ \hline
        0 & 20.13 (5.02) & 80.95 (4.75) & 71.78 (0.67) & 3.35 (0.16) \\
        30 & 17.45 (3.87) & 70.52 (6.00) & 68.34 (3.25) & 6.01 (3.13) \\
        45 & 20.78 (3.97) & 56.67 (6.35) & 67.83 (2.96) & 6.03 (2.89) \\
        60 & 23.55 (4.17) & 34.01 (11.41) & 63.71 (7.40) & 5.80 (3.23) \\
        90 & 20.15 (4.25) & 6.47 (7.33) & 63.09 (6.42) & 6.53 (3.26) \\
        180 & 11.26 (0.89) & -55.03 (1.14) & 55.08 (10.98) & 7.09 (4.55) \\
        \hline
    \end{tabular}
    
    \caption{Contract Setting: Average episodic rewards and the mean number of apples and diamonds collected (SD in parentheses).}
    \label{tab:cost_rewards}
\end{table}

We observe a qualitative difference in the strategies learned by the Sender, as they focus on extracting surplus from the Receiver and thus indirectly benefit from  the collection of apples.  Comparing these results with the signalling setting in Table~\ref{tab:nocost_rewards}, overall, there is an increase in the Sender's overall utility (at a cost to the Receiver). The contract offered depends on the alignment of the agents. When the two agents are well aligned ($\theta< 90)$ the Sender sends information to the Receiver but charges a high amount for it ($c\geq0.8$). Once $\theta>90$, the two agents are no longer reasonably aligned and the Sender shifts to a signalling policy with $p=0$ (i.e. it only sends action advice that is in its own interest, not the Receiver's interest). The contract also drops to $c=0$ since the Receiver quickly learns that the information provided by the Sender has no value. If the Receiver has better observability ($v=5$), then it is less reliant on the Sender, which again results in the Sender supplying better-quality information.  As seen in Figure~\ref{fig:heatmaps_highvis} (in App.~\ref{apx:gridworld}), when the agents are reasonably well aligned, the Sender's proposals promise useful information ($p=1$) while extracting more than half of the Receiver's rewards ($c=0.6$).

These results demonstrate that the Sender learns to use contracts strategically, aligning its actions with the Receiver’s interests when they are reasonably compatible, and thereby improving its episodic reward compared to the signalling setting. Addressing \textbf{RQ4}, the introduction of contracts generally reduces the Receiver’s accumulated reward. As in the signalling setting, the learned contract strategies remain sensitive to the Receiver's quality of observations and to the degree of reward misalignment between the two agents.

\section{Conclusion}
We study repeated interactions between an information-rich Sender agent and a decision-making Receiver agent with misaligned incentives. Through experiments in two different environments, we find that the Sender optimizes its cumulative rewards by learning effective signalling policies to influence the Receiver (\textbf{RQ1}). The Receiver learns to use its own partial observations along with the Sender's signals to navigate the environment more effectively, thus improving its cumulative rewards. These learned policies depend on the degree of alignment of their incentives and the quality of Receiver's observations (\textbf{RQ2}). Specifically, the Sender’s signalling strategies tend to align more closely with the Receiver’s interests when the Receiver’s observations are of higher quality, and conversely prioritize the Sender’s own interests when the Receiver’s observability is limited. In addition, we investigate the use of linear contracts that allow the Sender to price its information. Our results show that the Sender can leverage contracts to extract additional surplus (\textbf{RQ3}) and contract usage generally reduces the Receiver's total reward (\textbf{RQ4}).

Future work could explore extensions to more complex multi-agent environments such as social dilemmas where multiple agents act in the environment but information asymmetry still exists. More generally, given the increasing interest in agentic markets and human-AI collaboration, it is imperative to explore alternative mechanisms and contract designs that promote more balanced outcomes for the interacting agents to mitigate the exploitation of low-information participants.

\clearpage
\bibliographystyle{abbrv}
\bibliography{refs}
\clearpage

\appendix

\begin{center}
    {\Large \textbf{Appendix}}
\end{center}
\section{Modeling Choices}
\label{apx:model}
We study settings where there are two different agents, a \emph{Sender (S)} and a \emph{Receiver (R)}, interacting in some environment. There are some key asymmetries in the problem. Namely,
\begin{enumerate}
    \item The Sender (S) has an informational advantage over the Receiver (R) in that it can observe more of the environment.  S can decide whether or not to share this information with R.
    \item The Receiver (R) has agency, in that it can act in the environment. The Sender (S) must rely on the actions of the Receiver.
    \item Both R and S obtain rewards from the environment. However, the rewards may be different and may not be fully aligned.
\end{enumerate}

A single interaction of this kind has been studied extensively as Bayesian persuasion~\cite{kamenica_2011,kamenica_survey_2019}. We study the repeated interaction setting of the same problem with the possibility of payments (through contracts), but retain two key components of that model --- Sender's commitment and direct signalling. 

\textbf{Direct signalling} is standard in the Bayesian persuasion literature (also has parallels to action advising in MARL transfer learning~\cite{dasilva_2019_survey}), where the signal set is constrained to the Receiver’s action space ($A_R$). This is because expanding the signal space would not necessarily change the Receiver’s eventual best response. While Bayesian persuasion in general allows arbitrary signal spaces, it is without loss of generality to restrict attention to direct recommendation schemes, where the signal set equals the Receiver’s action space. This is a standard reduction in information design: any richer signal set can be merged into at most $|A_R|$ messages, since only the induced receiver action matters for equilibrium behavior. 

Within direct signaling, there are many possible families of signalling policies that the Sender can commit to, but we deliberately restrict ourselves to using a single scalar --- the \textbf{commitment probability}, $p$, \emph{i.e.}, the Sender's advise is sampled from the Receiver-optimal policy $\pi'_R$ with probability $p$. This makes the strategy space tractable for learning agents (through MAB) and the simplicity of a single parameter helps us better understand the high-level dynamics we seek to study. Note that this assumes Sender knows complete reward specification of the Receiver. In many real-world scenarios like the platform-seller scenario described in Section 1, this assumption is partly justified: platforms often observe aggregate outcomes, system dynamics, and seller responses at scale, and can infer incentive structures through repeated interactions, allowing them to approximate sellers’ reward functions.

Another key aspect of our model is the payments modeled through \textbf{linear contracts} that allow the Sender to extract additional surplus from the Receiver. Here again, we ground our work in the extensive economics and recent CS literature on contracts~\cite{duetting2024algorithmiccontracttheorysurvey,Salanie_2005}. We adopt simple linear contracts, where the contract is parametrized by $c$, the fraction of Receiver reward transferred to the Sender. As is common in the literature, the contract is proposed by one party (the Sender here), and the other party can only accept or reject it, \emph{i.e.}, it is a ``take it or leave it'' offer and there is \emph{no bargaining}. If the Receiver accepts the contract, Sender suggests action advise at every time step, and a fraction $c$ of Receiver's reward is transferred. If the Receiver rejects it, the Receiver is on its own for that episode, i.e., the Receiver acts based on its local observations only. We note that contract enforcement is not an issue in our setting because this is generally considered to be a modeling abstraction in the literature. The aim is to examine how learning agents behave when such alignment tools are available. 

The Sender moves first and deals with a single decision --- what contract to propose. The Receiver has two decisions at two different levels. At an episodic level, it has to decide whether the proposed contract should be accepted or not. Within the episode, at each timestep, the Receiver uses its local observation, the Sender's action advise (if available), and contract state to decide on the action to be executed in the environment. The exact sequence of events in an episode is clarified below.
\begin{enumerate}
    \item At the start of an episode, Sender proposes a contract $\langle p,c \rangle$. It learns to do this using MAB where each arm corresponds to a specific discretized contract.
    \item Receiver accepts or rejects it. This is learned through contextual bandits where the context is the contract, and the two arms correspond to accept and reject. 
    \item At any timestep within the episode, Receiver makes local observation. Sender sends action advise $a'$, which is drawn from Receiver-optimal policy $\pi'_R$ with probability $p$, and from $\pi'_S$ with probability $1-p$.
    \item Receiver uses its local observation, action advise $a'$, and the contract state to choose its action (using standard RL algorithms), which changes the environment. 
    \item Both Sender and Receiver get environment rewards due to the Receiver's action. Additionally, a fraction $c$ of the Receiver's reward is transferred to the Sender. 
    \item At the end of an episode, both Sender and Receiver use their episodic rewards to update the arm values for contract proposal and contract acceptance, respectively.
\end{enumerate}

\section{Optimal Contracts Analysis}
\label{apx:contracts}
We denote the episodic return of an agent following policy $\pi$ as $J(\pi)$, given by:
\[
J(\pi) = \mathbb{E}_{\tau \sim \pi} \left[ \sum_{t=0}^{T-1} \gamma^t r_{t+1} \right],
\]

In particular, we are interested in the utilities (average episodic reward) for both S and R, while following policies the hypothetical optimal policies from the perspectives of S and R, denoted by $\pi'_S$ and $\pi'_R$, respectively.  

\[
V_{RR} = J_R(\pi'_R), V_{SR} = J_S(\pi'_R),
\]
\[
V_{SS} = J_S(\pi'_S), V_{RS} = J_R(\pi'_S)
\]

The utility obtained by the decision-making Receiver $R$ on its own without any intervention by sender is denoted by $U^0_R$, which primarily depends on the observation quality of the Receiver.

For ease of analysis, we normalize all utilities such that the Receiver’s baseline utility, $U_R^0 = 0$. We use $\tilde{V}$ to denote such scaled utilities. For example, $\tilde{V}_{RR} = V_{RR} - U^0_R$.

Recall that $p$ is the commitment probability, \emph{i.e.}, the probability with which the Sender's action recommendation is sampled from Receiver-optimal policy $\pi'_R$ (it is sampled from the Sender-optimal policy $\pi'_S$ otherwise). $c$ is the fraction of reward that Receiver agrees to pay the Sender in exchange for the action advise.

The expected utilities of both agents are given by:
\begin{align}
U_S(p,c) &= p \, [\tilde{V}_{SR} + c\tilde{V}_{RR}] + (1-p) \, [\tilde{V}_{SS} + c\tilde{V}_{RS}], \\
U_R(p,c) &= (1-c) \, [p\tilde{V}_{RR} + (1-p)\tilde{V}_{RS}].
\end{align}

With $U_R^0 = 0$, the obedience constraint gives
\[
U_R(p,c) \ge 0  \implies p_{\min} = -\frac{\tilde{V}_{RS}}{\tilde{V}_{RR} - \tilde{V}_{RS}}.
\]

The Sender maximizes $U_S(p,c)$ subject to $p \ge p_{\min}$.  
Since $U_S(p,c)$ is linear in $p$, its slope with respect to $p$ determines the optimal signaling probability:
\[
\frac{\partial U_S}{\partial p} = (\tilde{V}_{SR}-\tilde{V}_{SS}) + c(\tilde{V}_{RR}-\tilde{V}_{RS}).
\]

The slope depends on $c$, and therefore there's a critical value, denoted by $\hat{c}$, given by:
\begin{align}
\hat{c} = \frac{\tilde{V}_{SS} - \tilde{V}_{SR}}{\tilde{V}_{RR} - \tilde{V}_{RS}}
\label{eq:c_hat}
\end{align}
 
Thus, the optimal signaling probability is:
\begin{equation}
p^* =
\begin{cases}
\max(\min(0, p_{\min}),1), & c < \hat{c},\\
1, & c > \hat{c}.
\end{cases}
\label{eq:optimal_pc_apx}
\end{equation}

%When $c < \hat{c}$, it is purely an information design problem. The Sender wants to commit to signaling strategies that satisfy the Receiver's constraint. At that point, $U_R$ is nearly zero and there is not much to extract through contracts. As $c$ increases beyond $\hat{c}$, the Sender’s utility becomes aligned with the Receiver’s, and it fully commits to Receiver-optimal advice ($p^* = 1$). 

When $c < \hat{c}$, transfers are weak and the Sender’s incentives remain misaligned with the Receiver’s. The Sender therefore chooses the lowest feasible probability $p_{min}$ that ensures Receiver's participation. Thus, the interaction reduces to the signalling setting where $U_R \sim 0$, and little surplus can be extracted. Conversely, when $c > \hat{c}$, the transfer share is sufficiently high that the Sender’s incentive becomes aligned with the Receiver’s, and the optimal strategy is then full revelation, $p^* = 1$.
%Since $p_{\min}$ is independent of $c$, changes in the sharing parameter affect only the Sender’s incentive to align, not the Receiver’s feasibility.

\section{Recommendation Letter Theory}
\label{apx:letter}
In the recommendation letter problem, there are two agents, a professor (Sender) and a recruiter (Receiver)~\cite{dughmi_survey_2017}. The professor is writing a recommendation letter for their student who is being recruited by the recruiter. The student is either a strong candidate or a weak candidate, and the student quality is known to the professor but not to the recruiter. The recommendation letter serves as a binary signal (recommend/don't recommend) from the professor to the recruiter. The recruiter receives a reward of +1 for hiring a strong student and -1 for hiring a weak student. If no student is hired, both agents receive a reward of 0. The professor receives a reward of +1 whenever their student is hired, regardless of quality.
This problem captures the challenges of asymmetric information and misaligned incentives. If the professor (Sender) truthfully reported student quality, the recruiter (Receiver) would only hire strong students. By recommending all strong students and randomly recommending weak students, the professor can increase their expected utility.

Formally, a student’s type is drawn from $\{S,W\}$ with prior $Pr(S) = p_0 \leq \tfrac12$

The professor chooses a signalling rule $\sigma:\{S,W\}\to \Delta(\{R,\bar R\})$, parameterized by
\[
p_1 = \Pr(R|S), \qquad p_2 = \Pr(R|W).
\]

The recruiter forms posteriors by Bayes’ rule:
\begin{align*}
    \Pr(S|R) &= \frac{p_1p_0}{p_1p_0+p_2(1-p_0)}, \\
    \Pr(W|R) &= \frac{p_2(1-p_0)}{p_1p_0+p_2(1-p_0)}.
\end{align*}

If the recruiter hires after receiving a message $R$, their expected utility is
\[
U_{rec}(R) = \frac{p_1p_0 - p_2(1-p_0)}{p_1p_0+p_2(1-p_0)}.
\]

If they do not hire, $U_{rec}(\bar R)=0$.  The obedience constraint requires
\[
U_{rec}(R)\geq 0. \tag{OC}
\]

The professor’s expected utility, given that the recruiter hires whenever they recommend, i.e., $R$ is sent, is
\[
U_{prof}(p_1,p_2) = p_1p_0 + p_2(1-p_0).
\]

\bigskip
\noindent \textbf{Optimization problem:}
\[
\max_{p_1,p_2 \in [0,1]} \; U_{prof}(p_1,p_2) 
\quad \text{s.t. } (OC).
\]

$U_{prof}$ is increasing in $p_1$ and decreasing in $p_2$, so the solution is:
\[
p_1^* = 1, \qquad 
p_2^* = \frac{p_0}{1-p_0}
\]

At $(p_1^*,p_2^*)$, $U_{rec}=0$. We assume the recruiter breaks the tie by hiring, and the professor’s payoff is
\[
U_{prof}^* = 2p_0.
\]

Consider a specific example, where $p_0 = 1/3$, i.e, a randomly drawn student is strong with probability $\tfrac13$. If the professor uses the optimal signalling policy above, i.e., recommend a weak student with one half probability, then the professor increases their expected utility to $\tfrac23$.

We get a similar result through our theoretical analysis in Section~\ref{sec:optimal_p}. The Receiver-optimal policy $\pi'_R$ is to hire when student is of type $S$ and to not hire otherwise. The Sender-optimal policy $\pi'_S$ is to hire always. Therefore, the expected utilities from an interaction (with prior $p_0=\tfrac13$) are: 

\[
V_{RR} = \frac{1}{3} , V_{SR} = \frac{1}{3}, V_{SS} = 1, V_{RS} = -\frac{1}{3}, \texttt{and } U_R^0 = 0
\]

To get the optimal commitment probability, $p^*$, plugging these values into~\eqref{eq:optimal_p}, we get: $p^* = \frac{1}{2}$. We note that this compact optimal solution is equivalent to the optimal signalling policy $\langle 1,\, 0.5 \rangle$ derived above. When the student is of type $S$, the optimal policies $\pi'_S$ and $\pi'_R$ coincide, both recommending to hire. Hence, although $p^*$ represents an equal-probability mixture over $\pi'_S$ and $\pi'_R$, the resulting recommendation is the same, i.e., $Pr(R \mid S) = 1$. In contrast, when the student is of type $W$, $\pi'_S$ recommends hiring while $\pi'_R$ does not, yielding $Pr(R \mid W) = \tfrac{1}{2}$.

Next, we calculate the optimal contracts for this setting. 

\[
V_{RR} = \frac{1}{3} , V_{SR} = \frac{1}{3}, V_{SS} = 1, V_{RS} = -\frac{1}{3}, \texttt{and } U_R^0 = 0
\]

Plugging these values into~\eqref{eq:c_hat} gives $\hat{c} = 1$. Therefore, the misalignment is high enough in this setting, that contracts cannot offset it. So, the optimal contract is to revert to the optimal signalling policy, $p^* = \frac{1}{2}$ while charging any $c < 1$, as there is very little to be extracted from the Receiver. 

\section{Additional Results for Rec Letter}
\label{apx:letter_res}
In this appendix, we include more results for the recommendation letter experiments (Section 3.1). Each trial consists of 200,000 interactions, and we define an episode to be 50 interactions with the Sender proposing the contract at the start of an episode. 

\begin{figure}[!h]
    \centering

    \begin{subfigure}{0.48\linewidth}
        \centering
        \includegraphics[width=\linewidth]{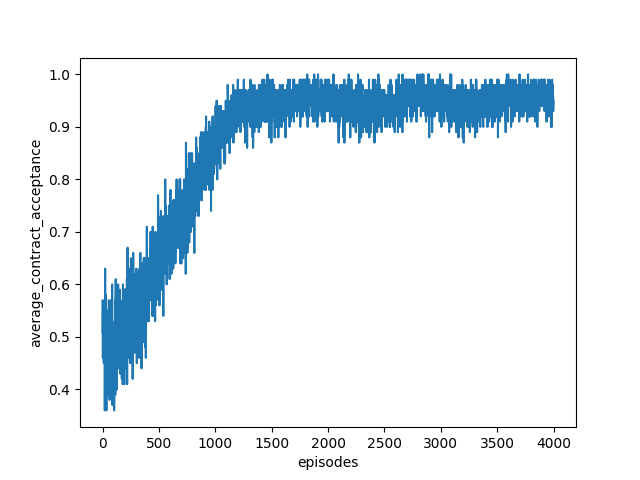}
        \caption{The average overall contract acceptance rates over 4000 episodes (200,000 interactions). These are averaged over 100 trials.}
        \label{fig:overall_contract_acceptance}
    \end{subfigure}
    \hfill
    \begin{subfigure}{0.48\linewidth}
        \centering
        \includegraphics[width=\linewidth]{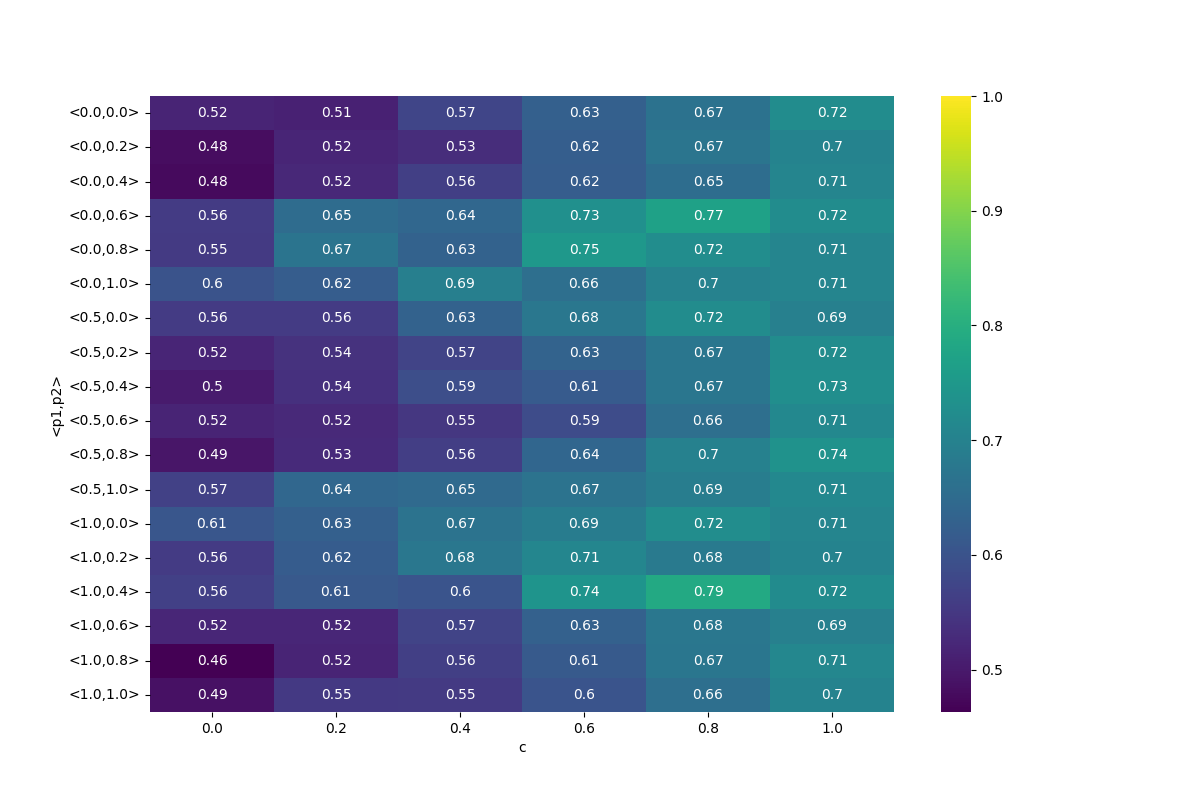}
        \caption{Contract setting: Average contract acceptance rates of each proposed contract averaged over 100 trials with each trial consisting of 4000 episodes.}
        \label{fig:contract_acceptance_rates}
    \end{subfigure}

\end{figure}

In the contract setting, where the Sender (professor) can charge a price, Figure~\ref{fig:overall_contract_acceptance} reports the Receiver’s overall contract acceptance rates, while Figure~\ref{fig:contract_acceptance_rates} breaks these down by individual contract. As learning progresses, the Receiver increasingly accepts the proposed contract, approaching near-universal acceptance. Acceptance rates are particularly high for the contracts that the Sender learns to propose most often, indicating that both agents’ learning processes have aligned to a stable equilibrium. Interestingly, the Receiver tends to accept costlier contracts more frequently than cheaper ones. We hypothesize that this pattern arises because higher contract costs $c \to 1$ may implicitly transfer more risk to the Sender, making such contracts relatively more appealing to the Receiver.

\begin{figure}[h]
    \centering
    \includegraphics[width=0.75\linewidth]{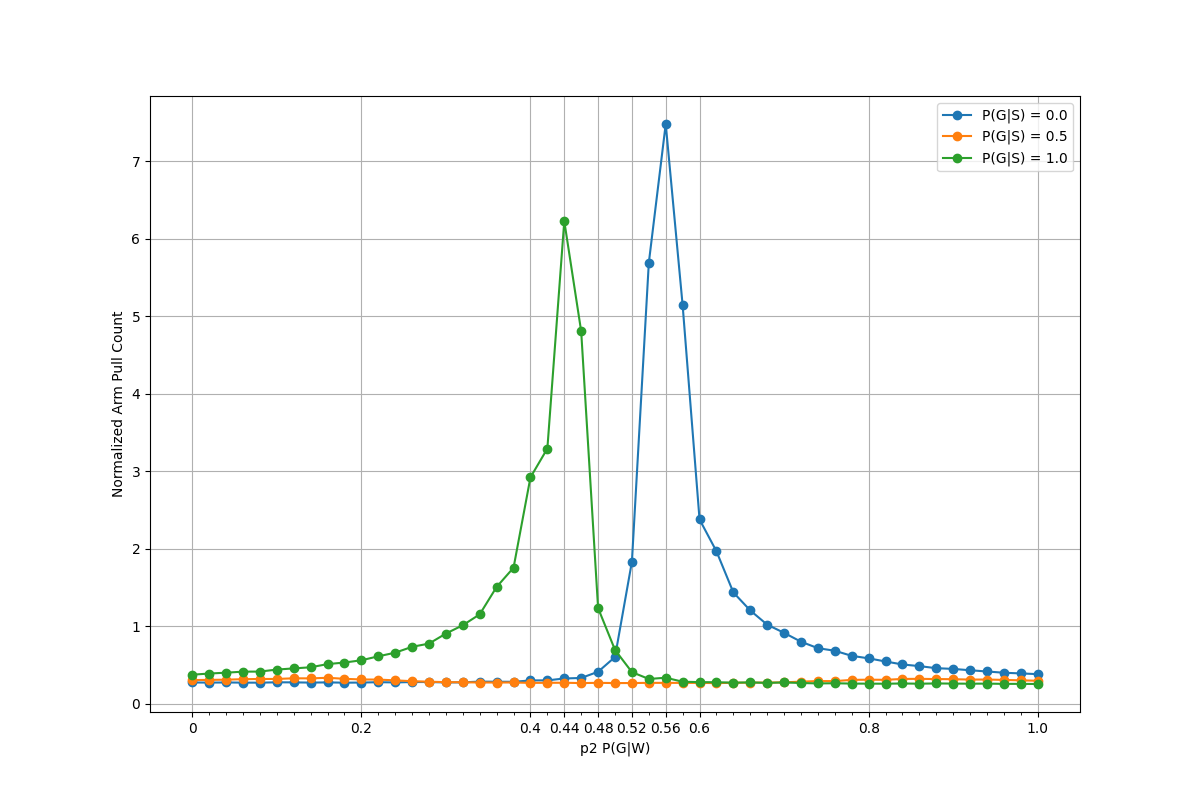}
    \caption{Average selection percentages of signalling strategies $\langle p_1, p_2 \rangle$ with $p_2$ discretized in 0.02 increments.}
    \label{fig:letter_heatmap_fullrange}
\end{figure}

In the signalling setting, where the Sender commits to a signalling strategy $\langle p_1, p_2 \rangle$, the main paper used a coarse discretization of $p_2$ in 0.05 increments. Here, we present results using a finer discretization of $p_2$ with 0.02 intervals. This effectively turns the Sender’s learning problem into a 153-arm bandit. The finer discretization increases exploration, slightly lowering the Sender’s average reward to 0.45 compared to the coarser setup. Figure~\ref{fig:letter_heatmap_fullrange} shows the average selection percentages for each signalling strategy. The learned strategy closely resembles what we observed with coarser discretization. Specifically, when $p_1 = 0$, the chosen $p_2$ values mostly fall between 0.38 and 0.48. When $p_1 = 1$, the selected $p_2$ values lie between 0.52 and 0.62, consistent with theoretical expectations.

\clearpage
\section{Additional Results for Gridworld}
\label{apx:gridworld}
\begin{figure*}[t]
    \centering
    \begin{subfigure}{\textwidth} % First subfigure (full width)
        \centering
        \includegraphics[width=\textwidth, height=0.37\pdfpageheight,keepaspectratio]{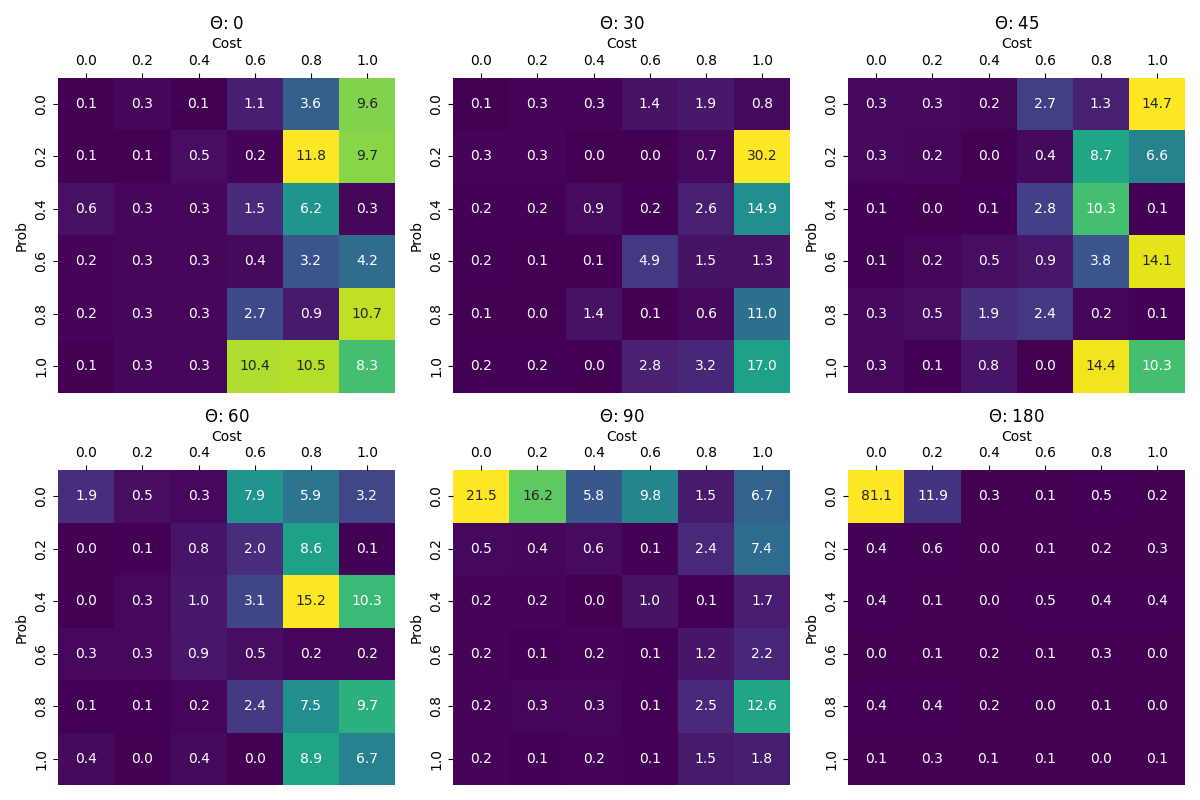}
        \caption{Low visibility setting $v=1$}
        \label{fig:heatmaps_lowvis}
    \end{subfigure}
    
    %\vspace{0.5cm} % Space between figures

    \begin{subfigure}{\textwidth} % Second subfigure (full width)
        \centering
        \includegraphics[width=\textwidth, height=0.37\pdfpageheight, keepaspectratio]{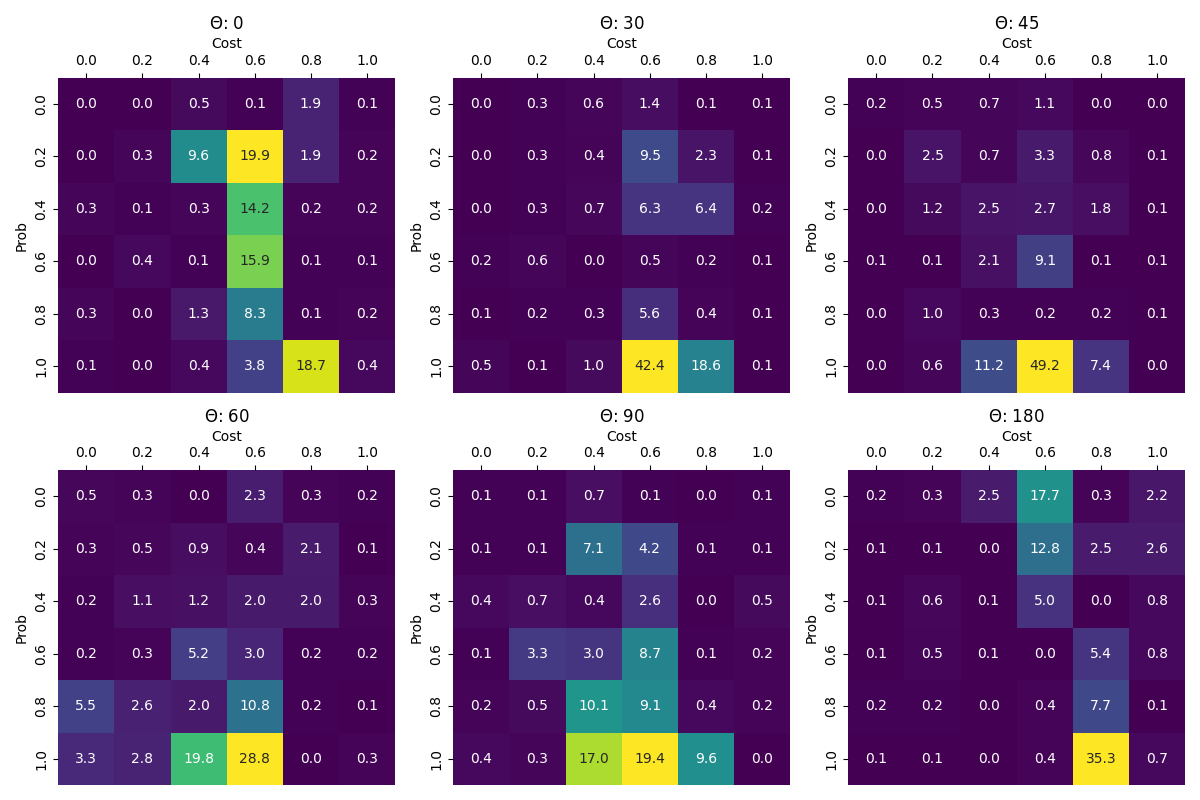}
        \caption{High visibility setting $v=5$}
        \label{fig:heatmaps_highvis}
    \end{subfigure}
    \caption{Contract strategies in gridworld: Heatmaps depicting average percentage of different contract proposals for different values of reward alignment $\theta$ and Receiver observability $v$.}
    %In each map, row values are the commitment probability $p$ while column values are the payment fraction $c$. Lighter colors indicate higher frequency.
    \label{fig:heatmaps}
\end{figure*}

\begin{table}[h]
   \centering
    \begin{tabular}{c|c}
        \toprule
       $\theta$ (in degrees) & $r_S$  \\ \hline
        0 & $\langle1,0\rangle $ \\
        30 & $\langle0.87,0.50\rangle $ \\
        45 & $\langle0.71,0.71\rangle $ \\
        60 & $\langle0.50,0.87\rangle $ \\
        90 & $\langle0,1\rangle $ \\
        180 & $\langle-1,0\rangle $    \\
        \bottomrule
    \end{tabular}
      \captionof{table}{The Sender reward vector $r_S$ for various values of $\theta$ while the Receiver reward vector $r_R$ is set as $<1,0>$.}
      \label{tab:reward_structure}
\end{table}

In this appendix, we present additional results for the GridWorld experiment using a stronger Receiver baseline. The Receiver now uses a breadth-first search (BFS) policy to locate the nearest apple, replacing the simple greedy heuristic from the main paper. Unlike the heuristic, BFS also explores the environment more systematically when no apple is visible. This change substantially improves the Receiver’s standalone performance: its average episodic reward increases from $-20.81$ to $-11.09$ under low visibility ($v=1$), and from $11.69$ to $22.33$ under high visibility ($v=5$). The following tables report the corresponding results for the signalling and contract settings, where the Sender interacts with this stronger Receiver. The values reported here are averages over 20 trials.

Overall, using a BFS-based Receiver reduces the Sender’s relative advantage compared to the greedy baseline, as the Receiver can locate apples more effectively on its own. The Sender still leverages contracts to extract surplus, but the improvement in Sender rewards (over the signalling-only setting) is less pronounced. When the Receiver is stronger, it tends to reject exploitative contracts and retain a larger share of the total reward. 

\begin{table}[!h]
%data_nocost_bfs = "logs_from_cc/log_nocost_1759427129.521622"
\centering
\scriptsize
\begin{tabular}{ccccc}
\hline
$\theta$ & Receiver & Sender & Apple & Diamond \\ \hline
\multicolumn{5}{c}{v = 1} \\
\hline
0   & 48.46 (3.32) & 48.46 (3.32) & 70.81 (3.16) & 4.29 (0.32) \\
30  & 46.02 (4.11) & 45.13 (2.83) & 67.87 (4.14) & 17.13 (6.74) \\
45  & 36.90 (5.56) & 44.08 (4.60) & 58.34 (5.78) & 34.90 (11.02) \\
60  & 11.86 (6.37) & 41.90 (4.88) & 33.56 (6.34) & 54.49 (7.64) \\
90  & 2.05 (5.43)  & 25.29 (14.72) & 24.53 (5.01) & 47.75 (13.97) \\
180 & -9.49 (2.61) & -35.89 (3.01) & 14.22 (2.52) & 33.50 (25.37) \\
\hline
\multicolumn{5}{c}{v = 5} \\
\hline
0   & 48.94 (6.06) & 48.94 (6.06) & 71.27 (5.76) & 4.16 (0.43) \\
30  & 47.05 (4.31) & 42.83 (3.95) & 69.15 (4.13) & 10.81 (3.66) \\
45  & 40.06 (7.18) & 33.64 (5.97) & 62.19 (6.91) & 17.24 (6.20) \\
60  & 37.91 (7.93) & 22.82 (7.41) & 60.12 (7.52) & 17.68 (6.33) \\
90  & 36.97 (4.54) & -2.69 (3.13) & 59.14 (4.37) & 19.44 (3.04) \\
180 & 24.19 (9.00) & -68.66 (7.75) & 47.43 (8.37) & 9.77 (7.14) \\
\hline
\end{tabular}
\caption{Signalling Setting: Average episodic rewards for both agents and the mean number of apples and diamonds collected per episode (standard deviation in parentheses).}
\label{tab:nocost_rewards_bfs}

\end{table}

\begin{table}[!h]
\centering
\scriptsize
\begin{tabular}{ccccc}
\hline
$\theta$ & Receiver & Sender & Apple & Diamond \\ \hline
\multicolumn{5}{c}{v = 1} \\
\hline
0   & 24.91 (3.03) & 59.85 (4.28) & 72.35 (3.07) & 4.64 (0.48) \\
30  & 22.73 (3.07) & 53.01 (3.42) & 66.89 (2.63) & 17.99 (3.42) \\
45  & 16.73 (3.53) & 44.72 (7.66) & 56.28 (5.32) & 29.54 (5.88) \\
60  & 9.69 (2.90)  & 41.18 (4.71) & 41.59 (3.78) & 45.82 (6.87) \\
90  & 1.60 (3.82)  & 16.93 (19.30) & 25.26 (7.98) & 39.07 (18.42) \\
180 & 1.19 (3.81)  & -43.75 (5.89) & 23.82 (6.92) & 41.39 (14.25) \\
\hline
\multicolumn{5}{c}{v = 5} \\
\hline
0   & 28.51 (3.52) & 55.22 (6.18) & 68.95 (4.44) & 4.22 (0.51) \\
30  & 26.06 (6.22) & 41.39 (11.54) & 61.00 (8.70) & 10.08 (3.84) \\
45  & 26.43 (3.42) & 35.37 (6.87) & 60.13 (4.44) & 13.66 (5.68) \\
60  & 24.38 (3.62) & 20.11 (5.61) & 57.39 (4.86) & 10.42 (5.16) \\
90  & 21.96 (5.85) & -7.56 (5.34) & 53.60 (6.87) & 11.31 (3.60) \\
180 & 20.24 (5.03) & -67.00 (5.77) & 47.09 (6.48) & 8.26 (3.75) \\
\hline
\end{tabular}

\caption{Contract Setting: Average episodic rewards for both agents and the mean number of apples and diamonds collected per episode (standard deviation in parentheses).}
\label{tab:cost_rewards_bfs}
\end{table}

\clearpage

\begin{figure*}[!h]
    \centering
    \begin{subfigure}[t]{0.48\textwidth}
        \centering
        \includegraphics[width=\textwidth]{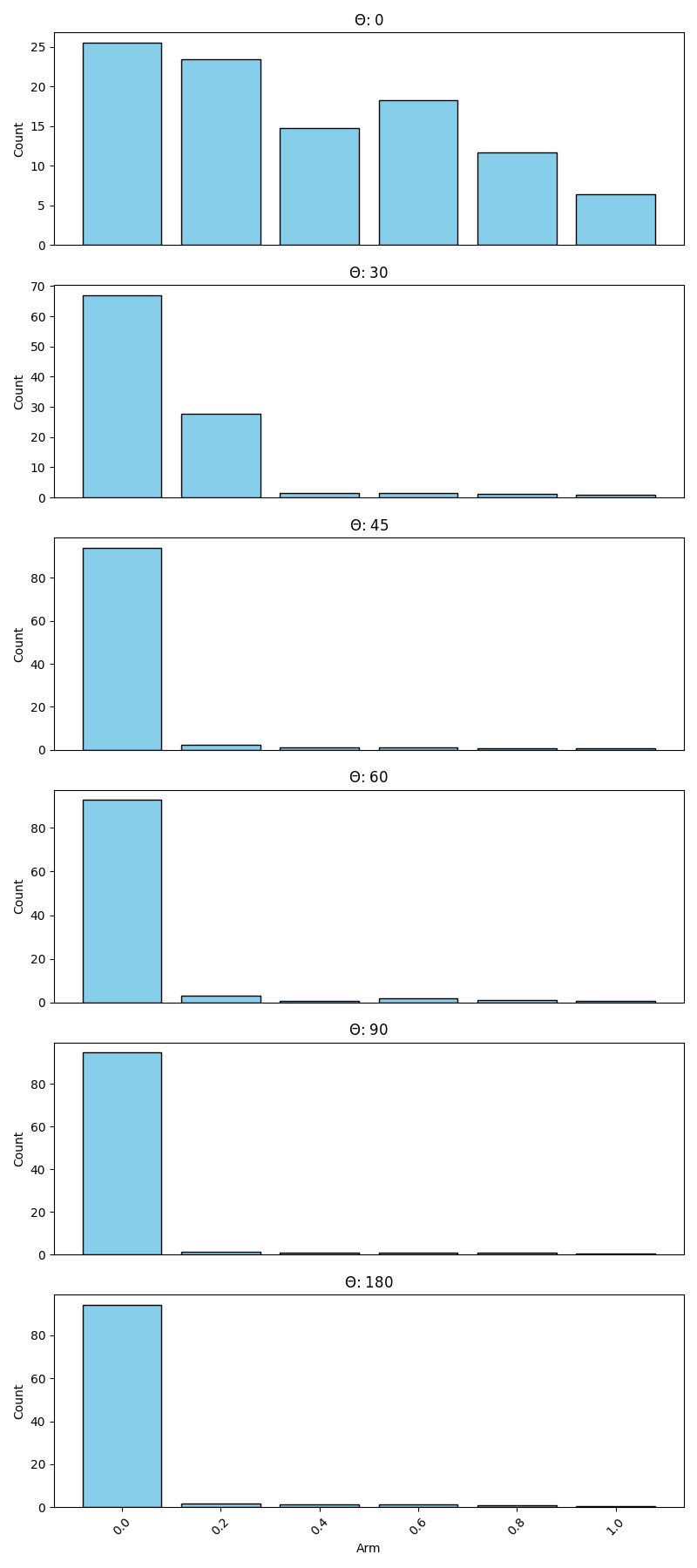} 
        \caption{Low visibility setting $v=1$}
        \label{fig:nocost_plots_lowvis}
    \end{subfigure}
    \hfill
    \begin{subfigure}[t]{0.48\textwidth}
        \centering
        \includegraphics[width=\textwidth]{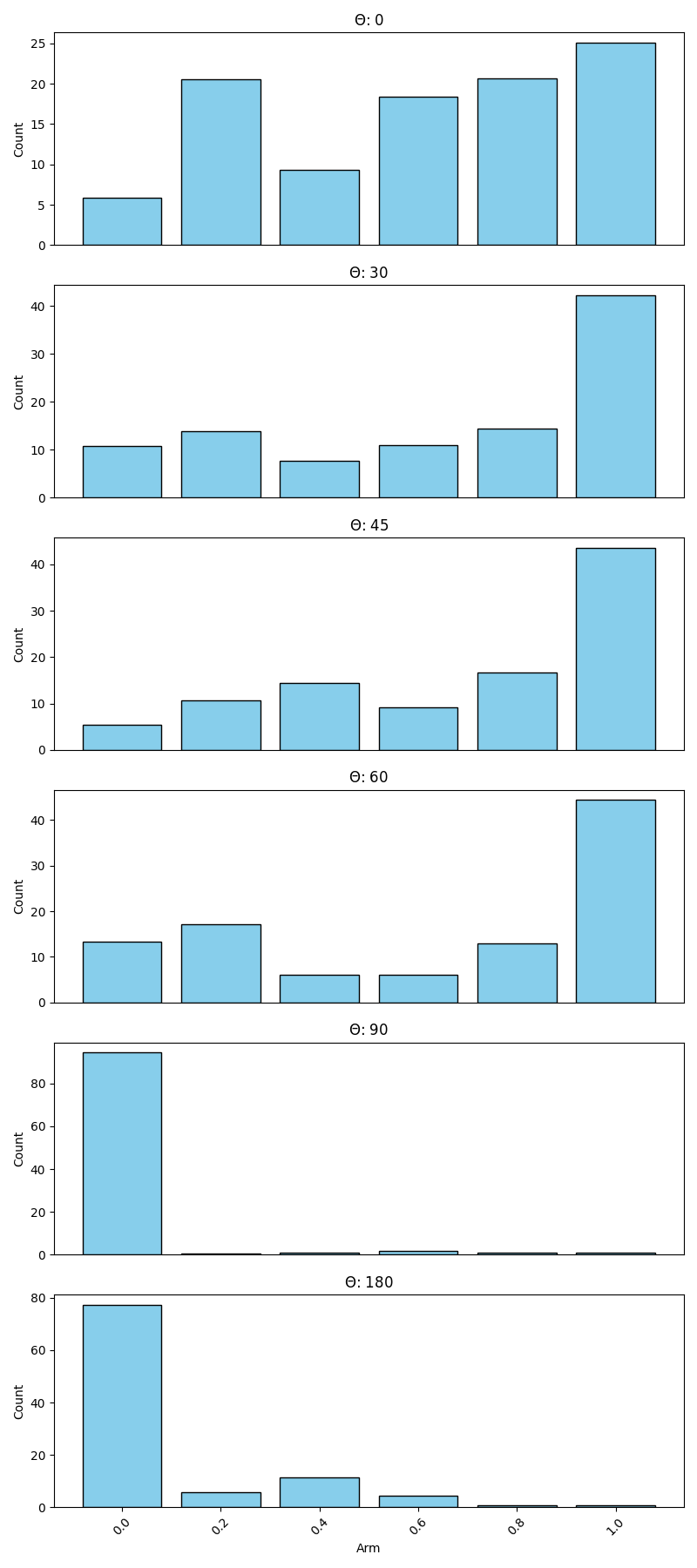} 
        \caption{High visibility setting $v=5$}
        \label{fig:nocost_plots_highvis}
    \end{subfigure}
    \caption{Alternate visualization of Figure~\ref{fig:compressed_arm_selection_nocost}. Signaling strategies in gridworld: Distribution of learned signalling strategies for the Sender in the no-cost information setting. Each bar represents the usage (as a percentage) of each signalling strategy.}
    \label{fig:nocost_plots}
\end{figure*}

\end{document}